\documentclass{ws-procs975x65}

\begin{document}

\title{Controlling Decoherence of Transported Quantum Spin Information in Semiconductor Spintronics}

\author{Branislav K. Nikoli\'c and Satofumi Souma}

\address{Department of Physics and Astronomy, Newark, DE 19716-2570, USA \\
E-mail: bnikolic@physics.udel.edu, WWW: http://www.physics.udel.edu/$\sim$bnikolic}

\maketitle

\abstracts{ We investigate quantum coherence of electron spin
transported through a  semiconductor spintronic device, where
spins are envisaged to be controlled by electrical means via
spin-orbit interactions. To quantify the degree of spin coherence,
which can be diminished  by an intrinsic mechanism where spin and
orbital degrees of freedom become entangled in the course of
transport involving spin-orbit interaction and scattering, we
study the decay of the off-diagonal elements of the spin density
matrix extracted directly from the Landauer transmission matrix of
quantum transport. This technique is applied to understand how to
preserve quantum interference effects of fragile superpositions of
spin states in ballistic and non-ballistic multichannel semiconductor
spintronic devices.}

\section{Introduction} \label{sec:introduction}

The major goal of recent vigorous efforts in spintronics is to
create, store, manipulate at a given location, and transport
coherent electron spin states through conventional semiconductor
heterostructures.\cite{spinreview} The two principal challenges
for new generation of spintronic devices are efficient injection
of spin into various semiconductor nanostructures and coherent
control of spin. In particular, preserving spin coherence, which
enables coherent superpositions of states $a|\!\! \uparrow \rangle + b |\!\! \downarrow \rangle$ and corresponding quantum-interference effects, is essential
for both quantum computing with spin-based qubits\cite{loss} and plethora of the proposed
classical information processing devices that encode information
into electron spin.\cite{spinreview,datta90}

The electrical control of spin via Rashba spin-orbit (SO)
interaction,\cite{rashba} which arises due to inversion asymmetry of
the confining electric potential for two-dimensional electron gas
(2DEG), has become highly influential concept in semiconductor
spintronics. A paradigmatic semiconductor spintronic device of this kind
is the Datta-Das spin-field-effect transistor\cite{datta90} (spin-FET)
where current passing through 2DEG in semiconductor
heterostructure is modulated by changing the strength of Rashba SO
interaction via gate electrode.\cite{nitta} The injected current can
be modulated in this scheme only if it is fully polarized, while precessing
spin has to remain {\em phase-coherent} during propagation between the two
ferromagnetic electrodes. Although spin injection into bulk semiconductors
has been demonstrated at low temperatures, creating and detecting
spin-polarized currents in high-mobility 2DEG has turned out to be a
much more demanding task.\cite{rashba}

For devices pushed into the  mesoscopic realm,\cite{datta_book} where at low temperature
$T \ll 1$K and at nanoscales full electron quantum state $|\Psi \rangle \in {\mathcal H}_o \otimes {\mathcal H}_s$ remains pure (in the tensor product of orbital and spin
Hilbert spaces) due to suppression of dephasing processes, it becomes possible to
modulate even unpolarized currents. In recently proposed spintronic ring device,\cite{ring} the conductance of unpolarized charge transport through a single channel ring
can be modulated between 0 and $2e^2/h$ by changing the Rashba electric field via
gate electrode covering the ring.\cite{nitta} This device exploits spin-dependent
quantum interference effects involving topological phases acquired in transport
through multiply-connected geometries, thereby avoiding ferromagnetic elements
and spin injection problems.
\begin{figure}[hp]
\centerline{\epsfxsize=1.6in \epsfbox{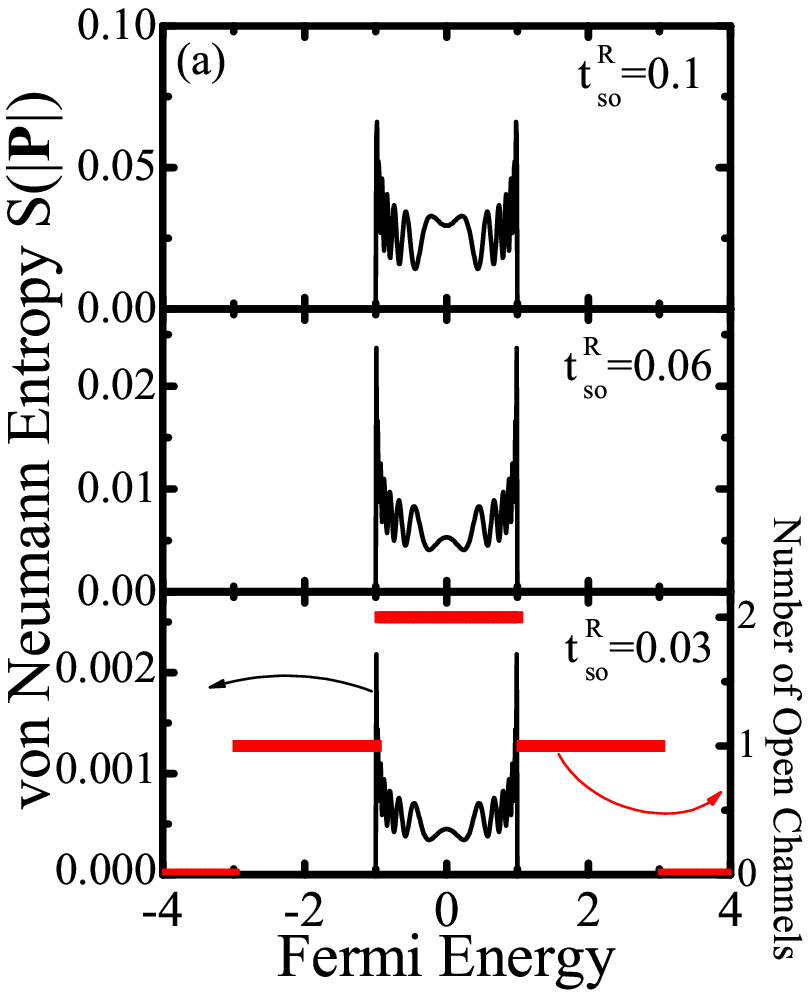} \epsfxsize=1.6in \epsfbox{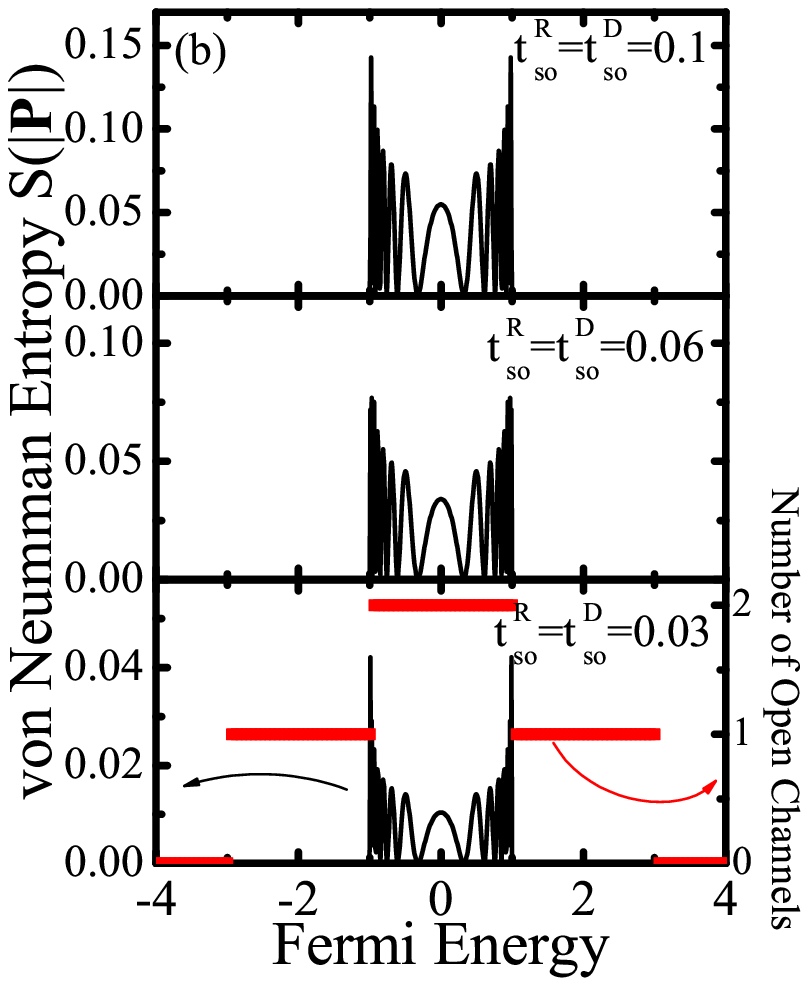} \epsfxsize=1.6in \epsfbox{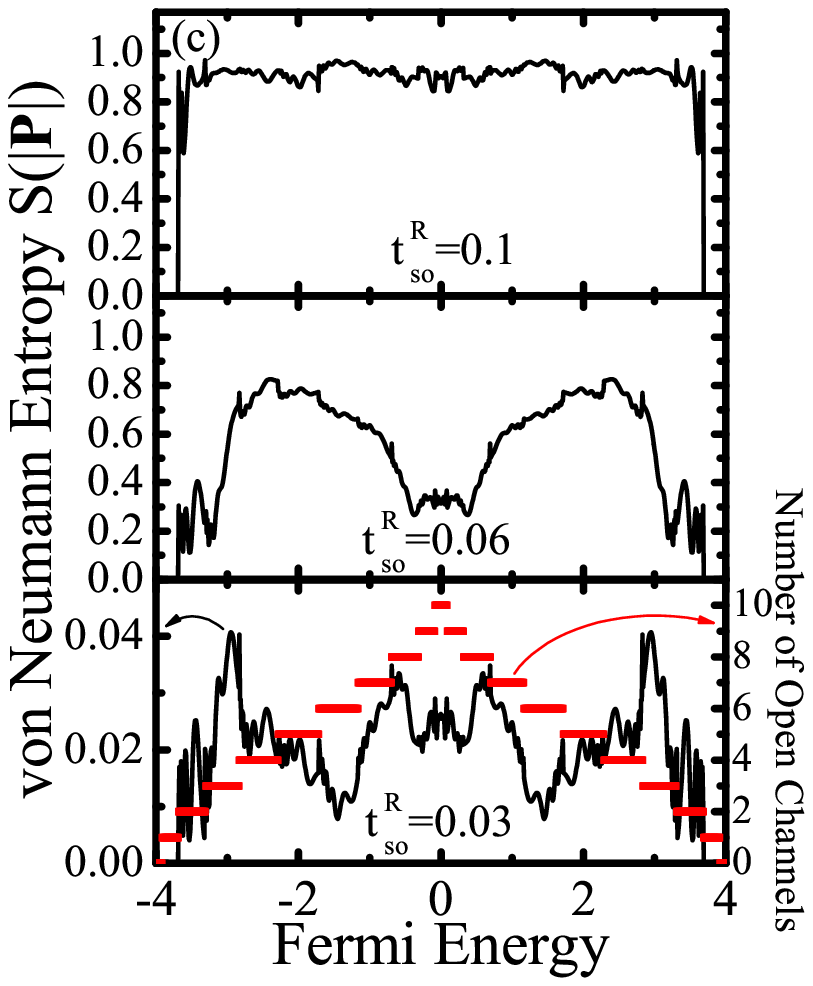}}
\caption{The von Neumann entropy of spins transmitted through a clean  semiconductor nanowire supporting maximum (around the band center) of two [panels (a) and (b)] or ten [panel (c)] conducting channels. The wire is modeled by a Hamiltonian Eq.~(\ref{eq:tbh}) with
different strengths of the SO couplings $t_{\rm so}^{\rm R}$ and $t_{\rm so}^{\rm D}$ on lattices: (a)  $2 \times 100$, (b) $2 \times 100$, and (c) $10 \times 100$. Note that
quantum coherence starts to decrease ($S >0$) when the second conducting channels becomes available for quantum transport. \label{fig:entropy}}
\end{figure}

However, even when all other spin decoherence mechanisms due to coupling to external
environment are suppressed,\cite{zurek} the same SO coupling that is envisaged to control
the spin can act to entangle spin and orbital quantum states. In such cases,  one
cannot associate a pure state $|\sigma \rangle \in {\mathcal H}_s$ to the spin degree of freedom any more.\cite{purity} The reduction of phase-coherence of an open spin quantum
system is formally described (as is the case of any decoherence process\cite{zurek}) as
the decrease of the off-diagonal elements of a two-level system density matrix
$\hat{\rho}_s = (1 + {\bf P} \cdot \hat{\bm{\sigma}})/2$, where ${\bf P}=(P_x,P_y,P_z)$ is
the {\em spin polarization vector}. The decoherence increases  the spin von Neumann
entropy  $S = -{\rm Tr} \, [\hat{\rho}_s \log_2 \hat{\rho}_s]$, which, in the case of spin-$\frac{1}{2}$ particle, is in one-to-one correspondence with the magnitude of the spin polarization vector $|{\bf P}|$:  $S(|{\bf P}|) =-(1+|{\bf P}|)/2 \log_2 ( 1+|{\bf P}|)/2 - (1-|{\bf P}|)/2 \log_2 (1-|{\bf P}|)/2$. For {\em pure} states, which are
fully coherent by definition, the polarization vector has unit magnitude $|{\bf P}|=1 \Leftrightarrow S=0$, while $|{\bf P}| = 0 \Leftrightarrow S = 1$  characterizes a non-pure state  that is completely unpolarized. For $0 < |{\bf P}| < 1$ ($0 < S(|{\bf P}| <1$), a spin-$\frac{1}{2}$ particle is in partially coherent quantum state which is described by
a {\em mixture} (or statistical superpositions) $\hat{\rho}_s^2 \neq \hat{\rho}_s$.

\section{Spin density matrix of detected current in semiconductor
nanostructures}

To understand the coherence properties  of transported spins (or mobile qubits\cite{mobile}),
we have developed a formalism\cite{purity} that extracts the spin density matrix from the
Landauer transmission matrix of quantum transport. The ${\bf t}$-matrix is traditionally employed\cite{purity} to compute  the spin resolved conductances $G^{\sigma' \sigma} =e^2/h \sum_{n'n} |{\bf t}_{n'n,\sigma' \sigma}|^2$. In the case of spin-dependent transport, the Landauer ${\bf t}$-matrix, which defines the outgoing asymptotic scattering state in the left right lead of a two-probe device $|{\rm out}\rangle$ when electron is injected in conducting channel $|n \rangle$ with spin $|\sigma \rangle$, also encodes the entanglement of orbital conducting channels (i.e., transverse propagating modes defined by the leads in the scattering picture of quantum transport\cite{datta_book}) and spin. This is due to the fact that  $|{\rm out} \rangle = \sum_{n',\sigma'} {\bf t}_{n'n,\sigma' \sigma} | n' \rangle \otimes |\sigma' \rangle$ is, in general, a non-separable state (i.e., a sum of the tensor product states $|n \rangle \otimes |\sigma \rangle$ that define spin-polarized conducting channels) because of spin-momentum entanglement\cite{peres} generated by spin-independent scattering (off lattice imperfections, phonons, nonmagnetic impurities, interfaces, ...) in the presence of SO interaction.\footnote{The entanglement of spin and orbital degrees of freedom\cite{purity} is somewhat different\cite{peres} from the familiar entanglement between different particles that can be widely separated and utilized for quantum communication,\cite{zurek,peres} because both degrees of freedom belong to the same particle. Formally similar entanglement of different degrees of freedom of one and the same particle has been pursued recently in Ref.\refcite{hekking}---an entanglement of a transverse wave function $|\Phi_n\rangle$ and a plane wave $| k  \rangle$ (with k-vector along the direction of transport) in the outgoing lead that form the basis of orbital conducting channels $|n \rangle = |\Phi_n\rangle \otimes |k \rangle$.}

By  viewing the current in the right lead of a two-probe spintronic device as an ensemble of {\em improper} mixtures, each  of which is generated after injecting electrons in different spin-polarized channels $|n \rangle \otimes |\sigma \rangle$ and propagating them through complicated  semiconductor environment, we introduce a {\em spin density matrix of the detected current}\cite{purity}
\begin{eqnarray} \label{eq:rhocurrent}
\hat{\rho}_c & = &  \frac{e^2/h}{n_\uparrow (G^{\uparrow \uparrow} + G^{\downarrow \uparrow}) + n_\downarrow(G^{\uparrow \downarrow} +
G^{\downarrow \downarrow})}  \\
&& \times \sum_{n^\prime,n=1}^M
\left( \begin{array}{cc}
     n_\uparrow |{\bf t}_{n^\prime n,\uparrow \uparrow}|^2 + n_\downarrow |{\bf t}_{n^\prime n,\uparrow \downarrow}|^2  &
       n_\uparrow {\bf t}_{n^\prime n, \uparrow \uparrow}
       {\bf t}^*_{n^\prime n,\downarrow \uparrow} + n_\downarrow {\bf t}_{n^\prime n,\uparrow \downarrow}
       {\bf t}^*_{n^\prime n,\downarrow \downarrow}  \\
         n_\uparrow {\bf t}^*_{n^\prime n,\uparrow \uparrow} {\bf
	t}_{n^\prime n,\downarrow \uparrow} + n_\downarrow {\bf t}^*_{n^\prime n,\uparrow \downarrow} {\bf t}_{n^\prime n,\downarrow \downarrow} &
       n_\uparrow |{\bf t}_{n^\prime n,\downarrow\uparrow}|^2 + n_\downarrow |{\bf t}_{n^\prime n,\downarrow \downarrow}|^2
  \end{array} \right). \nonumber
\end{eqnarray}
Here the injected current is assumed to be in the most general (i.e., partially polarized) state $\hat{\rho}_s  = n_\uparrow |\!\! \uparrow\rangle \langle \uparrow \!\!| + n_\downarrow |\!\!\downarrow \rangle \langle \downarrow \!\! |$. Special cases of injection of 100\%
spin-$\uparrow$ polarized or spin-$\downarrow$ polarized current correspond
to $n_\uparrow=1$, $n_\downarrow=0$ and $n_\uparrow=0$, $n_\downarrow=1$, respectively.
The spin polarization vector of the current is $(P_x,P_y,P_z) = {\rm Tr} \, [\hat{\rho}_c \hat{\bm{\sigma}}]$, and $\hat{\rho}_c$ also specifies the von Neumann entropy $S(|{\bf P}|)$ of the ensemble of transported spins that comprise the current. Thus, the equations
for $(P_x,P_y,P_z)$, together with the Landauer formula for spin-resolved charge conductances,\cite{purity} provide complete description of coupled spin-charge transport
in finite-size devices while intrinsically handling relevant boundary conditions.

We model generic semiconductor nanostructure by a single-particle Hamiltonian
\begin{eqnarray}\label{eq:tbh}
\hat{H} & = & \left( \sum_{\bf m} \varepsilon_{\bf m}|{\bf m} \rangle \langle {\bf m}|
  -  t \sum_{\langle {\bf m},{\bf m}^\prime \rangle} |{\bf m} \rangle \langle {\bf m}^\prime| \right) \otimes \hat{I}_s  + \frac{ \alpha  \hbar}{2a^2 t} (\hat{v}_y \otimes \hat{\sigma}_x - \hat{v}_x \otimes \hat{\sigma}_y) \nonumber \\
 && + \frac{\beta \hbar}{2a^2t}(\hat{v}_x \otimes \hat{\sigma}_x - \hat{v}_y
\otimes \hat{\sigma}_y),
\end{eqnarray}
written in the local-$s$-orbital$\otimes$spin basis on the lattice $M \times L$ (the
hopping $t$ between the orbitals sets the unit of energy), where  $(\hat{v}_x,\hat{v}_y, \hat{v}_z)$ is the velocity operator and we utilize the tensor product $\otimes$ of
operators in ${\mathcal H}_o \otimes {\mathcal H}_s$. The second term in the Hamiltonian is the Rashba SO interaction (whose electric field lies along the $z$-axis, while effective ${\bf k}$-dependent magnetic field ${\bf B} ({\bf k})$ of the SO interaction ${\bf B} ({\bf k}) \cdot \hat{\bm{\sigma}}$ emerges in the $xy$-plane), while the third term is the Dresselhaus one (arising from the bulk inversion asymmetry). In the ballistic wires of Sec.~\ref{sec:ballistic}, the on-site potential energy is $\varepsilon_{\bf m}=0$.  The
disorder in Sec.~\ref{sec:diffusive} is introduced through a standard random variable $\varepsilon_{\bf m} \in [-W/2,W/2]$.
\begin{figure}[hp]
\centerline{\epsfxsize=2.5in \epsfbox{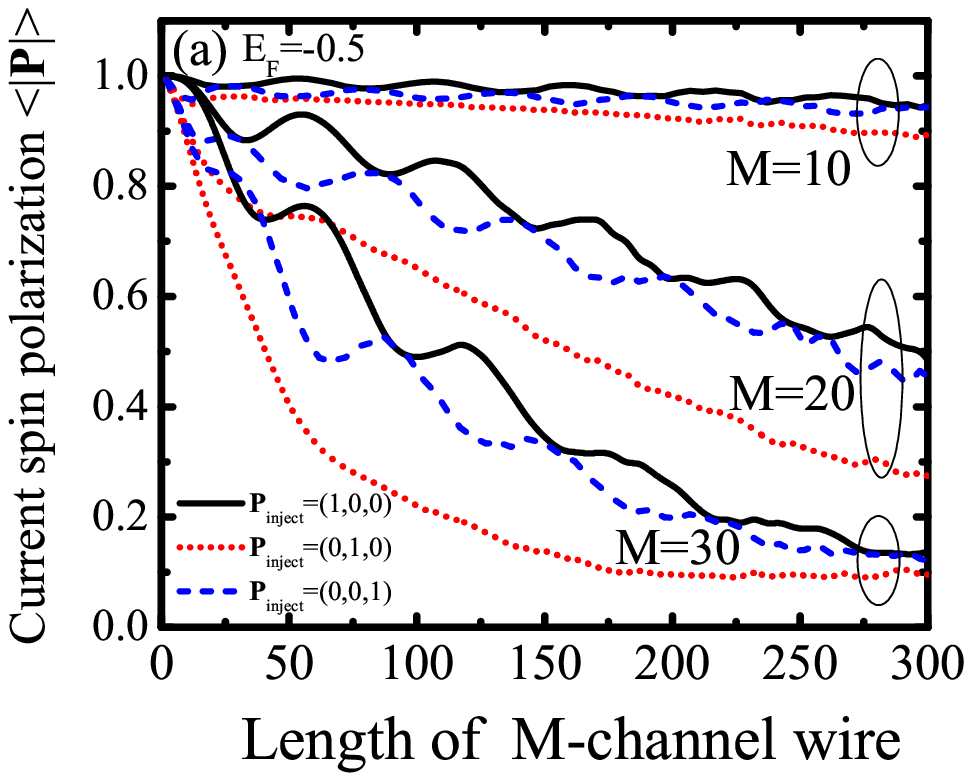} \epsfxsize=2.5in \epsfbox{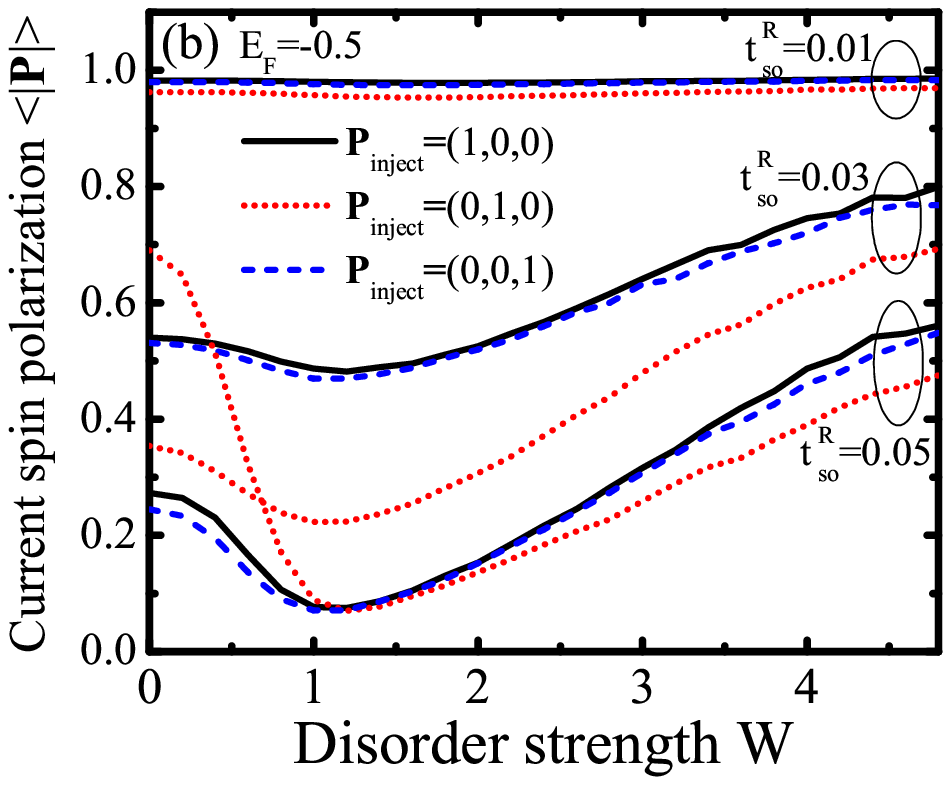}}
\caption{Quantum coherence of spins (injected with different direction
of ${\bf P}$) transmitted through  $M$-channel wires as a function of the
wire width [panel (a)] or the strength of the disorder $W$ and the Rashba SO
coupling $t_{\rm so}^{\rm R}$. The wires are modeled by the Hamiltonian in
Eq.~(\ref{eq:tbh}) on the lattice $M \times L$ [$\equiv 30 \times 100$ in panel (b)]. \label{fig:diffusive}}
\end{figure}
The spin-resolved transmission
matrix, ${\bf t} = 2 \sqrt{-\text{Im} \, \hat{\Sigma}_L^r \otimes
\hat{I}_s } \cdot \hat{G}^{r}_{1 N} \cdot  \sqrt{-\text{Im}\, \hat{\Sigma}_R^r \otimes \hat{I}_s}$ is obtained from the real$\otimes$spin space Green function $\hat{G}^{r}  =  [E \hat{I}_o \otimes \hat{I}_s- \hat{H} - \hat{\Sigma}^{r} \otimes \hat{I}_s]^{-1}$. Here $\hat{\Sigma}^{r}$ is the self-energy introduced by the leads,\cite{datta_book} and
$\hat{I}_o$ and $\hat{I}_s$ are the unit operators in ${\mathcal H}_o$ and ${\mathcal H}_s$,  respectively. The matrix elements of the SO terms in Eq.~(\ref{eq:tbh}) contain "SO hopping parameters"  $t_{\rm so}^{\rm R}=\alpha/2a$ and $t_{\rm so}^{\rm D}=\beta/2a$ that set the energy scales of the Rashba and Dresselhaus coupling, respectively.\footnote{In current experiments\cite{nitta} maximum achieved values of $t_{\rm so}^{\rm R}$ are of the order of $\sim 0.01t$.}

\section{Spin coherence in ballistic spin-FET-type devices} \label{sec:ballistic}

Despite advances in nanofabrication technology, it is still a challenge to fabricate semiconductor nanowire that contains only one transverse propagating mode. We investigate spin coherence  in multichannel {\em clean} wires in Figure~\ref{fig:entropy}, which plots the spin entropy $S(|{\bf P}|)$ as a function of the Fermi energy $E_F$ of electrons whose transmission matrix ${\bf t} (E_F)$ determines coupled spin-charge transport in a wire supporting at most two or ten orbital conducting channels.
\begin{figure}[hp]
\centerline{\epsfxsize=2.7in \epsfbox{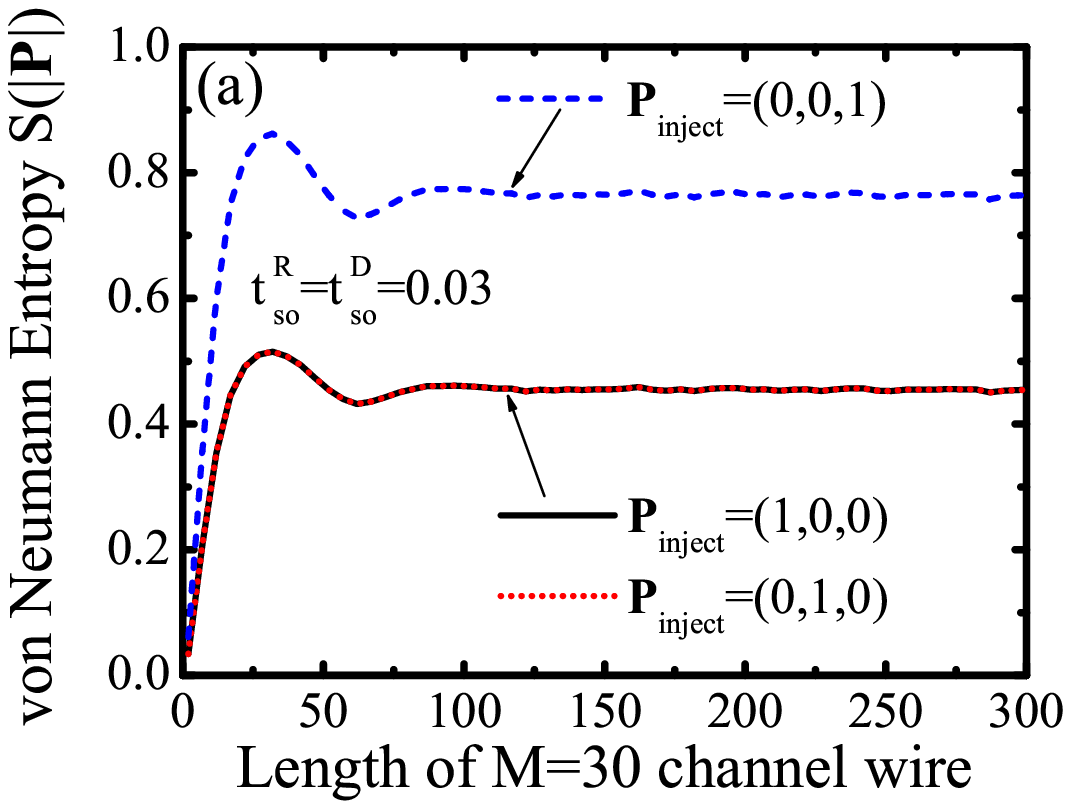} \epsfxsize=1.7in \epsfbox{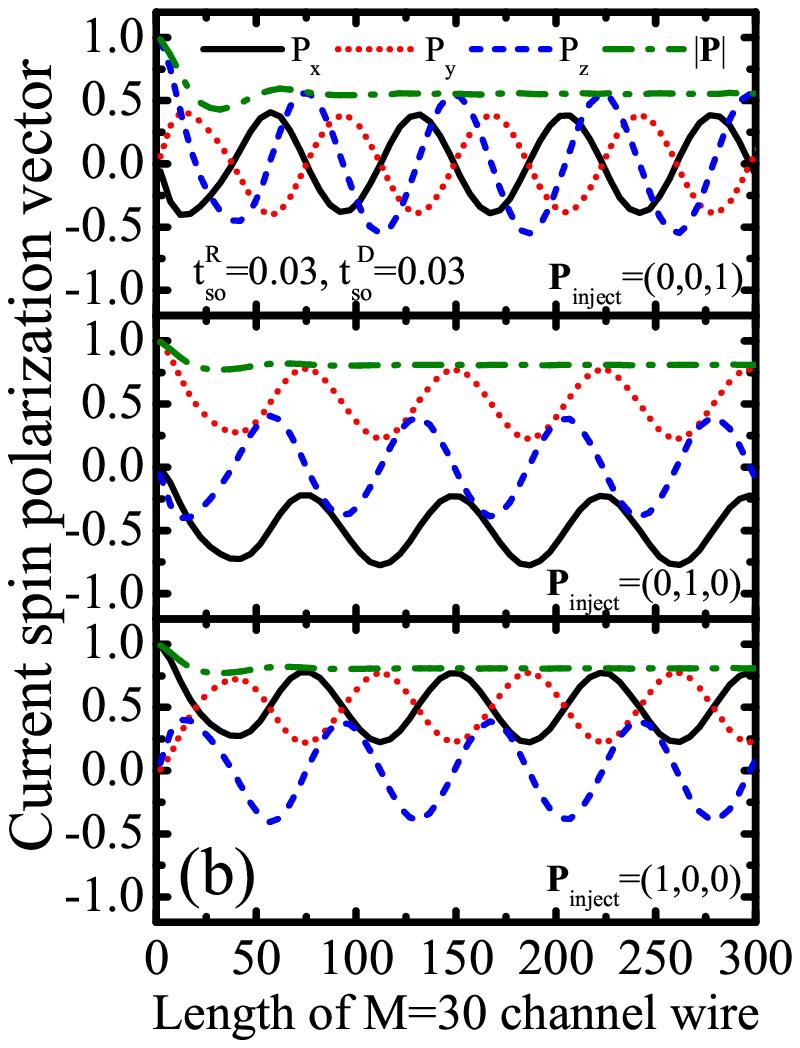}}
\caption{Spin coherence along the 2DEG region, modeled on the lattice $30 \times L$,
of a non-ballistic spin-FET device$^{14}$ (where $t_{\rm so}^{\rm R} = t_{\rm so}^{\rm D}$) for different directions of polarization of injected current. \label{fig:nonbal}}
\end{figure}
The current injected from the left lead is assumed to be fully polarized along the
direction of transport (the $x$-axis chosen here), as in the spin-FET operation where
such setup ensures high level of current modulation.\cite{datta90}  As long as only
one conducting channel is open, spin remains coherent since outgoing state in the
right lead is $ (a |\!\! \uparrow \rangle + b |\! \! \downarrow \rangle) \otimes |n=1 \rangle$. At exactly the same $E_F$ where the second channels $|n = 2\rangle$ becomes available for transport, $S(|{\bf P}|)$ becomes non-zero signaling that spin state loses its purity. This is due to the fact that at this $E_F$, the quantum state of transported spin becomes {\em entangled} to its orbital state,\cite{purity} $|{\rm out} \rangle = a |\!\! \nearrow \rangle \otimes |e_1 \rangle +  b |\!\! \swarrow \rangle \otimes |e_2 \rangle$. The scattering at the lead-semiconductor interface, which in the presence of the SO interaction generates such non-separable (i.e.,  entangled) outgoing state is induced by different nature of electron states in the wire and in the leads. While recent studies\cite{strong_rashba} have pointed out that for {\em strong} Rashba SO interaction this effect can lead to complete  suppression of spin injection, here we unearth how even moderate Rashba coupling in wires of few nanometers width can affect coherence of ballistically transported spins. This becomes increasingly detrimental when more channels are opened, as demonstrated in Fig.~\ref{fig:entropy} for a $M=10$ channel nanowire.

\section{Spin coherence in non-ballistic spin-FET-type devices} \label{sec:diffusive}

When transported charge scatters of spin-independent impurities in 2DEG with SO interaction, its spinor will get randomized and the  disorder-averaged current spin polarization $\langle |{\bf P}| \rangle$ will decay along the wire. This is as an old problem initiated by seminal (semiclassical) study of D'yakonov' and Perel (DP).\cite{dyakonov} We demonstrate the decay of spin coherence  within our quantum transport formalism in Fig.~\ref{fig:diffusive}, where the decay rate decreases in narrow wires thereby suppressing the DP mechanism.\cite{wl_spin} The effect acquires a transparent physical interpretation within the same framework invoked in Sec.~\ref{sec:ballistic}---the spin decoherence is facilitated when there are more conducting channels to which spin can entangle in the process of spin-independent charge scattering that induces transitions between the channels. Moreover, we find in Figure~\ref{fig:diffusive} quantum corrections to spin diffusion in strongly disordered systems, which capture Rashba spin precession beyond the DP theory or weak localization corrections to it (that are applicable for weak SO correction in random potential that can be treated perturbatively\cite{wl_spin}). The current spin polarization $\langle |{\bf P}| \rangle$ in the wires of fixed length is recovering with disorder as soon as the diffusive transport regime is entered. Within the picture of spin entangled to an effectively zero-temperature ``environment'' composed of orbital transport channels, this effect has a simple explanation for {\em arbitrary} disorder strength: as the disorder is increased,
some of the channels become closed for transport thereby reducing the number of
degenerate ``environmental'' quantum states that can entangle to spin.

Recently a lot of theoretical interest\cite{nonballistic} has been directed toward
relaxing  strict ballistic transport regime required in the original\cite{datta90}
spin-FET. In non-ballistic spin-FET proposal, tuning of equal Rashba and Dresselhaus SO interaction  $\alpha = \beta$ is predicted to cancell the spin randomization due to charge scattering. We confirm in Fig.~\ref{fig:nonbal} that spin coherence computed within our formalism indeed does not decay along such 2DEG wire. However, length-independent constant value of $\langle |{\bf P}| \rangle$ is set below one $(S > 0)$ and, moreover, depends on the spin-polarization properties of injected current. Thus, transported in such specially crafted 2DEG environment will remain in partially coherent state, rather than being described by a single spin wave function. This can be traced to the observation of diminished spin coherence in clean systems [see see panel (b) of  Fig.~\ref{fig:entropy}].

\section{Conclusions}

In conclusion, we have shown that transported electron spin in multichannel semiconductor spintronic devices (such as spin-FET\cite{datta90} or spintronic rings\cite{nitta,mring}) can
be subjected to the loss of coherence due to an interplay of SO interactions and any type of charge scattering (including the one at the lead-2DEG interface). Nonetheless, spins remain in
a partially coherent~\cite{partial_fano} state that can exhibit quantum interference
effects of reduced visibility\cite{hekking} [as indicated in panel (b) of Fig.~\ref{fig:nonbal}].

\section*{Acknowledgments}
We are grateful to F. E. Meijer, L. P. Z\^ arbo, and S. Welack for
valuable discussions.

\end{document}